# Predicting synthetic rescues in metabolic networks

Adilson E. Motter[1*], Natali Gulbahce[2], Eivind Almaas[3] and Albert-László Barabási[4]

[1]Department of Physics and Astronomy and Northwestern Institute on Complex Systems, Northwestern University, IL 60208, USA.

[2]Theoretical Division and Center for Nonlinear Studies, Los Alamos National Laboratory, NM 87545, USA. Center for Complex Network Research and Department of Physics, Northeastern University, Boston, MA 02115, USA. Center for Cancer Systems Biology, Dana Farber Cancer Institute, One Jimmy Fund Way, Boston, MA 02115, USA.

[3]Network Biology and Microbial Systems Group, Biosciences and Biotechnology Division, Lawrence Livermore National Laboratory, CA 94550, USA.

[4]Center for Complex Network Research and Departments of Physics and Computer Science, University of Notre Dame, Notre Dame, IN 46556, USA. Center for Complex Network Research and Departments of Physics, Biology and Computer Science, Northeastern University, Boston, MA 02115, USA.

[*] Corresponding author. E-mail: motter@northwestern.edu. Tel/Fax: 1-847-491-4611/9982

**An important goal of medical research is to develop methods to recover the loss of cellular function due to mutations and other defects. Many approaches based on gene therapy aim to repair the defective gene or to insert genes with compensatory function. Here, we propose an alternative, network-based strategy that aims to restore biological function by forcing the cell to either bypass the functions affected by the defective gene, or to compensate for the lost function. Focusing on the metabolism of single-cell organisms, we computationally study mutants that lack an essential enzyme, and thus are unable to grow or have a significantly reduced growth rate. We show that several of these mutants can be turned into viable organisms through additional gene deletions that restore their growth rate. In a rather counterintuitive fashion, this is achieved via additional damage to the metabolic network. Using flux balance-based approaches, we identify a number of synthetically viable gene pairs, in which the removal of one enzyme-encoding gene results in a nonviable phenotype, while the deletion of a second enzyme-encoding gene rescues the organism. The systematic network-based identification of compensatory rescue effects may open new avenues for genetic interventions.**

## Introduction

Recent advances in systems and network biology indicate that specific cellular functions are rarely carried out by single genes, but rather by groups of cellular components, including genes, proteins, and metabolites (Elena and Lenski, 1997; Hartwell *et al*, 1999; Vogelstein *et al*, 2000; Barabási and Oltvai, 2004; Bonhoeffer *et al*, 2004; Albert 2005; Segrè *et al*, 2005). Such a network-based view changes the way we think about the impact of mutations and other genetic defects: the damage caused by a malfunctioning protein or gene is often not localized, but spreads through the cellular network, leading to a loss of cellular function by incapacitating one or several functional modules (Goh *et al*, 2007; Barabási 2007). The increasingly sophisticated experimental tools that help us systematically map various cellular interactions offer hope that in the future we will be able to focus not only on the individual components, but also monitor and explore the global changes in the cellular network induced by the defective gene or protein. Such network-based approaches indicate that the loss of proteins involved in a large number of protein-protein interactions often results in the death of the organism, a finding that may be useful for the design of antibiotics or cancer drugs. Yet for most genetic diseases, particularly those caused by germline mutations, the goal is not to kill the cell, but to recover the lost cellular function or limit the existing damage. This raises an important question: can we develop network-based strategies to predict how to recover function that may have been lost due to defective genes?

In single-cell organisms, the frequently observed reduction in an organism's growth rate following a gene deletion often represents only a transient effect, reflecting the fact that the metabolic network of the mutant operates in a suboptimal regime until appropriate regulatory changes and mutations accumulate to bring the metabolic system to a new optimal steady state (Fong and Palsson, 2004; Herring *et al*, 2006). Experiments in fixed nutrient environments show that after many generations mutants typically increase their growth rate, converging through adaptation to a new optimal value predicted by Flux Balance Analysis (FBA) (Edwards and Palsson, 2000). If the growth rate in this optimal state is zero, then the organism cannot grow, indicating that the deleted gene is essential. We will refer to these genes as *optimally essential*.

Often experiments observe no growth for mutants missing a metabolic enzyme that are predicted to be viable by FBA, prompting us to classify the deleted gene as essential. One potential explanation for the observed discrepancy is that the gene may have an unknown function, regulatory or other, whose absence inhibits growth. Yet, for some enzymes an equally compelling explanation is the following: an important challenge of each mutant is to reproduce until the evolutionary tuning of its regulatory system approaches the new optimal growth state. Thus even if FBA predicts a nonzero optimal growth rate, some mutants may not survive due to their inability to grow in the suboptimal state right after the gene deletion (Fig. 1(a), red line). The growth rate of the organisms shortly after a gene deletion can be effectively calculated using the minimization of metabolic flux adjustment (MOMA) method (Segrè *et al*, 2002), a variant of FBA. In the following we will call a gene *suboptimally essential* if the optimal (FBA predicted) growth rate is nonzero in its absence, while the MOMA predicted growth rate is zero. Therefore, experiments will probably classify organisms missing a suboptimally essential gene as unable to grow. However, in contrast with the optimally essential genes, a mutant missing a suboptimally essential gene would be determined as viable if its metabolism and regulatory system had the chance to re-adjust to its environment.

Here we show that the growth rate of an organism lacking a suboptimally essential gene may be restored via the removal of other enzyme-encoding genes. We will refer to this as the *Lazarus effect*, as it restores the growth of mutants initially classified as nonviable by experiments since they displayed zero growth rates. We also discuss *suboptimal recovery*, a weaker manifestation of the proposed mechanism, which forces viable mutants to increase their growth rate following additional gene deletions. Our approach is inspired by a method proposed in (Motter, 2004) to control cascading failures in complex networks and by microbial optimization methods for the targeted production of metabolites (Burgard *et al*, 2003; Pharkya and Maranas, 2006).

## Results

The principle underlying the proposed rescue effect is illustrated schematically in Fig. 1(b-e). Consider the situation where, in the wild-type organism, the optimal growth state corresponds to the utilization of the $M_1 \rightarrow M_2 \rightarrow M_4$ pathway, i.e., the flux of reactions involving the $M_3$ metabolite is either zero or close to zero. In the early state after the deletion of the enzyme catalyzing the $M_2 \rightarrow M_4$ reaction, metabolism operates suboptimally (Fig. 1(c)) by minimizing the necessary flux rearrangement compared to the optimal wild-type flux state (Fig. 1(b)). The optimal post-deletion state, however, requires more drastic flux reorganization, sending most of the flux through the $M_1 \rightarrow M_3 \rightarrow M_4$ pathway (Fig. 1(d)). It would take additional regulatory and metabolic adjustments to reach this new optimal state. This process can be facilitated by deleting the enzyme catalyzing the $M_1 \rightarrow M_2$ reaction, forcing the cell to use the optimal $M_1 \rightarrow M_3 \rightarrow M_4$ pathway (Fig. 1(e)). Therefore, by suppressing stoichiometrically inefficient pathways, we can force the cell to enhance the activity of a more efficient set of reactions, resulting in an increased growth rate. Our goal is to show that such additional deletions, whose role is to enhance the activity of the most efficient pathways, can be predicted by systematically comparing the suboptimal and the optimal fluxes under the same conditions.

To implement the approach described in Fig. 1, we developed an algorithm to identify rescue deletions for all mutants missing an enzyme-encoding gene. For this we use MOMA to determine the suboptimal fluxes $\mathbf{v}^{1}_{MOMA}$ characterizing the mutant shortly after a gene deletion (Fig. 1(c)) and FBA to predict a flux state $\mathbf{v}^{1}_{FBA}$ compatible with optimal growth for the mutant (Fig. 1(d)). If the mutant's metabolism operates suboptimally after the gene deletion, the FBA predicted growth rate for the mutant is larger than the MOMA predicted growth rate, and thus, we have a chance to intervene and increase the suboptimal growth rate. In this case, based on the difference in flux pattern between $\mathbf{v}^{1}_{FBA}$ and $\mathbf{v}^{1}_{MOMA}$ (see Methods), we test a set of secondary rescue gene deletions that aim to reduce the difference between the suboptimal and the optimal growth rate (Fig. 1(e)) by using MOMA to determine the new metabolic flux state $\mathbf{v}^2_{MOMA}$. If appropriate rescue gene deletions are identified, the obtained growth rate $G^2_{MOMA}$ is higher than the growth rate of the original mutant, concluding our procedure. If the MOMA predicted growth rate for the original mutant is zero, the rescue deletions can bring along the *Lazarus effect*, inducing a nonzero growth rate; if it is nonzero, the rescue deletions may induce a *suboptimal recovery*, increasing the mutant's growth rate towards its optimal FBA predicted value. Note that the identified rescue deletions do not change the optimal growth rate, but affect only the suboptimal growth rate (see Methods). The new terminology related to this recovery mechanism is summarized in Table I.

We illustrate the proposed procedure in Fig. 2 for the TCA cycle of *E. coli* MG1655 fed arabinose as the sole carbon source (see Methods and Supplementary Information 1). MOMA predicts that the deletion of the *fbaA* gene rearranges the fluxes throughout the whole cycle and inhibits the production of phenylalanine, tyrosine, and L-lysine (dotted reactions in Fig. 2(b)), which represent necessary building blocks of the biomass (cf. Fig. 2(a)). Thus, the suboptimal growth rate of this mutant is zero, a prediction supported by experiments in arabinose media (Fraenkel, 1987). In contrast, FBA indicates that a nonzero growth rate can be achieved by a global rearrangement of the flux states (Fig 2c), resulting in changes in flux magnitudes and directions (e.g. the *sucCD* reaction). Consequently, the organism *could* grow if it could get past its suboptimal state when, soon after the gene deletion, its growth rate is zero. We can force the organism to approach the new optimal state by deleting, for example, the genes *aceA* and *sucAB*, which catalyze reactions that are active in the suboptimal state (Fig. 2(b)) but are not active in the optimal state (Fig 2(c)). These two rescue deletions will activate the production of *all* biomass components after rerouting the fluxes through the pentose phosphate pathway (Fig. 2(d)), and result in a nonzero growth rate, rescuing the otherwise nonviable mutant.

The growth rate of the *fbaA* mutant can be further enhanced by deleting additional genes that catalyze reactions that are inactive in the optimal state (see Methods). We illustrate this in Fig. 3(a), which shows the predicted suboptimal growth rate of the *fbaA*-deficient *E. coli* mutant after the concurrent removal of several genes in addition to *aceA* and *sucAB*. While the rescue deletion of *aceA* is sufficient to recover growth, the additional deletion of *sucAB*, *tnaB*, *xapB*, and *prr* further enhances the growth rate, with a large enhancement predicted after the removal of *tnaB*. The biomass production reaches a plateau of about 67% of the wild-type biomass production rate after the deletion of forty genes. The situation is similar for suboptimal recovery: as we show in Fig. 3(b) for the case of the *nuoA* mutant with glucose as the carbon source, additional gene deletions can increase the growth rate of the mutant, eventually approaching 59% of its wild-type optimal value.

Systematically applying our method to the *E. coli* metabolism in glucose minimal medium, we identified 6 suboptimally essential genes, which represent candidates for the Lazarus effect, and 17 candidates for suboptimal recovery (see Fig. 4(a)). Most of the mutants miss genes involved in the central metabolism, while a few miss genes that participate in amino acid metabolism and transport processes. Of particular interest are mutants with the genes *pfk*, *fbaA* or *tpiA* deleted, whose essentiality has been tested and is supported by experiments (Fraenkel, 1987). As we show in Table SI (Supplementary Information 1) and Fig. 4(a), the growth rate of these mutants is restored by additional targeted gene deletions that increase the suboptimal growth rate from zero to more than 45% of the wild-type growth rate.

In Fig. 4(b) we show that, for various media, the increase in the biomass production rate obtained after the deletion of a *single* rescue gene can be more than 10% of the wild-type rate. In other cases, however, we need to simultaneously delete several genes to rescue growth. This is illustrated in Fig. 3(b), where we show that the growth performance of nonviable *tpiA*-deficient mutants in a glucose medium can be restored only through the concurrent deletion of six genes, *aceA*, *gadA*, *gadB*, *lpdA, tynA* and *gpt*, representing a six-viable set, which is the converse of the k-robust set necessary to suppress cellular growth (Deutscher *et al*, 2006). The suboptimal *tpiA*-mutant uses the glyoxylate pathway, which is shut down by these rescue deletions. Our prediction, that the glyoxylate pathway is not needed in the optimal state, is supported by a recent experimental observation (Fong *et al*, 2006). This observation indicates that the flux of the glyoxylate pathway in viable but not fully evolved *tpiA*-mutants is initially nonzero. However, over the course of a few weeks of adaptive evolution in glucose media, the glyoxylate flux converges to zero (Fong *et al*, 2006). Once the six genes are absent, the concurrent deletion of additional genes can further increase the organism's growth rate (Fig. 3(b)).

Note that, while the proposed rescue procedure works in all media, the list of mutants that can be rescued by additional deletions as well as the necessary rescue deletions depends on the tested medium. Indeed, we find that the number of *E. coli* mutants whose growth rate increases by more than 10% of the wild-type growth rate after rescue deletions is 8, 21 and 25 in minimal acetate, minimal glucose, and rich media, respectively. Therefore, the rescue effect is more frequent in richer media, where the increased availability of substrates in the environment increases the number of non-essential metabolic genes that can be deleted to improve performance. Furthermore, the proposed rescue mechanism is expected to work for all organisms, allowing us to predict rescue deletions each time an accurate metabolic reconstruction is available. To show this, we determined all single-gene rescues that can recover the growth rate by more than 1% of the wild-type rate in glucose media for deletion mutants of three reconstructed organisms with very different genomes: *H. pylori* (341 enzyme-encoding genes), *E. coli* (660), and *S. cerevisiae* (750). Interestingly, the obtained number of mutant-rescue combinations for these organisms, 58, 94, and 58, respectively, is consistently large and to some extent comparable despite the significant differences in their metabolism.

In our analysis of the most significant cases for the eukaryote *S. cerevisiae*, we predict the Lazarus effect for 3 mutants and suboptimal recovery for 11 other gene deletions in a glucose minimal medium (Fig. 4(c)). It is interesting to note that several of these genes are human orthologs (Steinmetz *et al*, 2002; BiGG, 2007), including genes *pfk*, *tpi1*, *lpd1*, and *mir1*. Of the three mutants predicted to exhibit the Lazarus effect, two of them have been experimentally verified to be nonviable, while positive growth has been observed in experiments for the third one (SGD, 2007). The observed small disagreement with our predictions is probably due to the incompleteness of the reconstructed model or the fact that the organisms in the experiments were not fully adapted to the medium modeled in our computations. As in the case of *E. coli*, the intensity of the recovery generally increases with the number of genes in the rescue set and best recovery may involve up to 50 genes in the examples shown in Fig. 4(c). However, we predict that a comparable recovery can be obtained with significantly fewer deletions (Supplementary Information 2 and 3). In particular, we also find numerous examples in several media of single-gene rescue deletions resulting in a significant increase of biomass production in *S. cerevisiae* mutants (Fig. 4(d)).

The focus so far in this work has been on developing an approach to computationally predict synthetic rescue in metabolic networks. Importantly, the founding hypothesis of this approach, that the suboptimal growth rate of an organism can be improved by the removal of properly selected genes, is consistent with experiments. To demonstrate this, in Fig. 5 we reanalyze experimental results (Fong and Palsson, 2004; Fischer and Sauer, 2005) for the growth rate of several mutants in their suboptimal state, before and after a gene deletion. The compiled data in Fig. 5(a) indicates that the suboptimal growth rates of *E. coli* MG1655 can indeed improve considerably after the deletion of selected enzyme-encoding genes, an effect observed in multiple environments. In Fig. 5(b) we show similar results for *B. subtilis* 168, following the removal of genes involved in various cellular functions. Note that, the metabolism of the wild-type strains in these experiments is not fully adapted to the media and operates in a suboptimal regime. This experimental evidence, together with the power of FBA (Edwards *et al*, 2001; Ibarra *et al*, 2002) and MOMA (Segrè *et al*, 2002; Shlomi *et al*, 2005) to predict the optimal and suboptimal growth rate of an organism in agreement with experimental data, supports our hypothesis that properly selected gene deletions can improve the growth rate of an organism that has not yet adapted to its environment. To further substantiate this claim, we calculated the reaction fluxes determined by experimental uptake and growth rates (Fong and Palsson, 2004) as well as the corresponding optimal reaction fluxes. We used these flux distributions to test our assumption that gene deletions increasing (not increasing) growth tend to be associated with reactions whose fluxes are much larger (smaller) than the optimal fluxes. As shown in Table SIV (Supplementary Information 1), a total of 20 out of 22 *E. coli* mutants analyzed are correctly predicted with this assumption, in support of the proposed rescue mechanism.

## Discussion

The mechanism behind the rescue effect introduced above does not depend on the specific details of MOMA or FBA; in fact, any computational or experimental methodology that can help us estimate the metabolic fluxes can be used to identify candidates for rescue deletions. For example, one could use $^{13}C$-tracer techniques (Sauer, 2004) to experimentally determine the reaction fluxes of the suboptimal gene-deficient strain and an optimal, or close to optimal, version of the same strain. Candidates for rescue deletions typically correspond to genes catalyzing reactions that are active in the suboptimal state but inactive in the optimal state. By identifying these reactions experimentally, one could minimize biases due to inaccurate modeling in the identification of the candidate rescue deletions. However, we find that our *in silico* predictions are robust to parameter choices and do not rely on the fine-tuning of metabolic fluxes or environmental conditions (see Supplementary Information 1). Furthermore, we predict that the rescue set in an impaired cell is not unique, and the number of rescue combinations that lead to the same effect generally increases with the number of genes in the set (Supplementary Information 4 and 5). These observations corroborate the feasibility of systematic experimental implementation of synthetic rescues. Indeed, the main difficulties expected in verifying our predictions, namely the inaccuracies in matching real genetic and environmental conditions as well as potential side effects of rescue deletions due to, e.g., unknown function, are substantially alleviated by the robustness and flexibility of the rescue interactions. This generality, which transcends particular computational methods, could serve as a bridge to implementations of our approach in multi-cellular organisms, as it facilitates the control of undesirable effects in the recovery of specific cellular functions.

The possibility of rescuing a mutant using additional gene deletions is a general mechanism not limited to metabolism. For example, the removal of *comA* and *sigD* genes enhances the growth rate of *B. subtilis* (see Fig. 5), despite the fact that they have no known enzymatic functions. Additionally, it has been observed in *E. coli* that *edd*-deficient mutants grow at a reduced rate and *eda*-deficient mutants do not grow at all in a gluconate medium, while the double *edd*/*eda* mutant is viable. In this case the mechanism for the rescue effect is different from the one discussed above: upon the deletion of *eda*, the cells accumulate toxic compounds; this accumulation stops when *edd* is also deleted (Fraenkel, 1987). Gene-deletion induced rescue processes have been observed previously in mammalian cells as well. For example, Irs2 knockout mice develop diabetes in 6 to 8 weeks (Kushner *et al*, 2004; Hahnfeldt and Hlatky, 2005). Yet, the additional knockout of Ptp1b partially compensates for the lack of Irs2, doubling the survival time. Similar effects were documented for mutations in HK1.ros, HK1.fas and HK1.TGK-$\alpha$ in tandem with the loss of *p53* gene (Wang *et al*, 2000; Hahnfeldt and Hlatky, 2005). The mechanisms behind these examples involve mostly local gene-gene interactions, as opposed to the global effect we have systematically unveiled here. They indicate, however, that organisms could be characterized in general by potentially extensive sets of *synthetically viable* double knockouts, representing gene pairs for which the double mutant is viable while one of the single mutants is not. High throughput techniques, increasingly used to identify synthetically lethal pairs (Tong *et al*, 2001; Ooi *et al*, 2003), could be used to uncover such synthetically viable gene pairs as well. Other techniques may be developed to identify similar interactions between gene *sets*, as proposed above. The results could be used to detect new genetic compensatory mechanisms and would offer a better understanding of cellular functions, just as synthetically lethal gene pairs have deepened our understanding of genetic interactions (Wong *et al*, 2004; Bonne *et al*, 2007). Furthermore, our results force us to adjust the current paradigm of gene essentiality: even if the deletion of a gene is lethal, the gene is not necessarily essential to support life (Kobayashi *et al*, 2003; Pál *et al*, 2006; Glass *et al*, 2006; Hashimoto *et al*, 2005) because the organism's ability to metabolize biomass may be restored by additional gene deletions.

Finally, our findings may also offer a new alternative to restore the loss of cellular function caused by specific mutations. Indeed, current approaches based on gene therapy (Ho and Commins, 2001; Kimmelman 2005; Kaiser 2005) may trigger abnormal activity associated with the vector and insertion site, such as oncogenesis, or reinforce the activity of pathways encoding malfunctioning products of the faulty gene, such as misfolding proteins. From a drug design and therapy perspective, it may be more advantageous to block the activity of selected pathways rather than trying to restore the activity of a faulty gene or protein. Specific previous experimental studies that can be related to the recovery mechanism reported here corroborate the feasibility of such an approach. It has been observed, for example, that *Myc* deletions rescues the *Apc* deficiency in murine small intestine. This presumably takes place because *Myc* is required for gene activation involved in cancer development often following *Apc* innactivation (Sansom *et al*, 2007). In a different study, the combination of antibiotics exhibiting hyper-antagonistic interactions, where the combined effect of two antibiotics is weaker than at least one alone, has been shown to select against resistant strains (Chait *et al*, 2007). In the context of our work this means that, in the two-drug sublethal medium, the "deficient" bacterial cells (non-resistant strain) prevail. Another example is found in studies of *E. coli* mutants unable to grow anaerobically on glucose and other hexoses when gene *adh* (ethanol production) or gene *pta* (acetic acid production) is inactivated, but the mutant with both genes deactivated will grow through the production of lactic acid as the major fermentation product (Gupta and Clark, 1989). On the other hand, a non-fermenting mutant of *E. coli*, NZN111, is rescued to ferment glucose through the inactivation of the *ptsG* gene, resulting in the production of succinate, acetate and ethanol by rerouting fluxes that would go through the partially blocked pathways of pyruvate in NZN111 (Chatterjee *et al*, 2001). In addition, it has been shown that the concurrent deletion of genes *zwf*, *sfcA*, *maeB*, *ndh*, *ldhA*, and *frdA* maximizes the biomass yield in wild-type *E. coli* MG1655 by eliminating the elementary metabolic modes associated with low biomass yield (Trinh *et al*, 2006). The latter study also demonstrates the feasibility of creating mutants with several targeted rescue deletions (Causey *et al*, 2003), in support of our suggestions. These examples are not limited to a single organism and can be interpreted as different manifestations of a common rescue mechanism. Therefore, a combination of experimental and computational studies aimed at systematically uncovering synthetically viable gene pairs and gene sets, as well as the underlying rescue effects, may open new avenues for the next generation of therapeutic strategies.

## Methods

### Constraint-based approach

For a network with *m* metabolites and *n* reactions, the stoichiometric constraints are represented by $\sum_j S_{ij}\,\nu_j = 0$, where $S=(S_{ij})$ is the $m\times n$ matrix of stoichiometric coefficients, and $\mathbf{v}=(\nu_j)$ is the vector of fluxes. The individual fluxes are limited by thermodynamic constraints, substrate availability, and the maximum reaction rates supported by the catalyzing enzymes and transporting proteins, as

$$\alpha_j \leq \nu_j \leq \beta_j, \qquad (1)$$

where $\alpha_j = \beta_j = 0$ for uptake reactions of substrates not available in the medium. The biomass production is incorporated as an additional reaction $\sum_i c_i\, x_i \xrightarrow{G} 1$ *biomass unit*, where the stoichiometric coefficient $c_i$ corresponds to the experimentally measured biomass composition of metabolite $x_i$ (Edwards and Palsson, 2000). FBA consists of finding a metabolic state that satisfies these constraints while maximizing the biomass flux G. The deletion of genes responsible for the production of the enzymes involved in reaction *j* corresponds to imposing the bounds $\alpha_j = \beta_j = 0$ in Eq. (1). MOMA aims to find a solution $\mathbf{v}_{MOMA}$, compatible with the constraints imposed to the mutant, while being closest to the original metabolic state $\mathbf{v}_{FBA}$ in terms of Euclidean distance in the space of fluxes. Our implementations of FBA and MOMA are based on the optimization softwares GNU Linear Programming Kit (Makhorin, 2001) and Object-Oriented Quadratic Programming Package (Gertz and Wright, 2001), respectively, and have been tested using independent implementations of the CPLEX solver (ILOG CPLEX).

### Identifying rescue gene deletions

Consider a strain generated by the deletion of a metabolic gene that constrains at least one of the nonzero metabolic fluxes of the wild-type organism and such that the biomass flux after this deletion is $G^1_{MOMA} < G^1_{FBA}$. To increase $G^1_{MOMA}$, we compute the vector of all metabolic fluxes $\mathbf{v}^1_{FBA} = (\nu_j{}^1_{FBA})$ predicted by FBA and use them to define a second gene deletion. This deletion is defined by identifying the minimum number of metabolic genes that deactivate most or all reactions *j* with $\nu_j{}^1_{FBA} = 0$. These gene deletions force the metabolic system to operate closer to the optimal regime predicted by FBA, while they do not change the FBA fluxes and the predicted steady state biomass production (i.e. $G^2_{FBA} = G^1_{FBA}$). The corresponding changes in the MOMA predicted fluxes are expected to increase the biomass flux from $G^1_{MOMA}$ to $G^2_{MOMA} > G^1_{MOMA}$. We recursively discard the deletions that have no impact on the MOMA predicted biomass production. The number of gene deletions is further reduced through recursively activating genes from the rescue set that contribute the least to the increase in $G^2_{MOMA}$. Note that this approach increases the biomass production itself, which is not necessarily related to the biomass yield considered in metabolic engineering studies (Causey *et al*, 2003; Trinh *et al*, 2006).

Algorithmically, we start with a mutant strain defined by the deletion of one or more metabolic genes, and identify the sets of rescue deletions by adhering to the following procedure:

1. Calculate the FBA optimal flux vector $\mathbf{v}^{wt}_{FBA}$ for the wild-type strain.
2. Calculate the FBA optimal flux vector $\mathbf{v}^{1}_{FBA}$ for the mutant strain.
3. Calculate the MOMA flux vector $\mathbf{v}^{1}_{MOMA}$ for the mutant strain, using $\mathbf{v}^{wt}_{FBA}$ as a reference flux.
4. Continue if biomass flux $G^1_{MOMA} < G^1_{FBA}$, as a set of rescue deletions may exist.
5. Identify reaction set K consisting of all reactions *j* such that $v_j{}^1{}_{MOMA} \neq 0$ and $v_j{}^1{}_{FBA} = 0$. Potential recovery is implemented by setting $(\alpha_j, \beta_j) = (0,0)$ for every $j \in K$.
6. Identify K* as self-consistent subset of K by obeying the gene-enzyme relationships.
7. Incrementally reduce K* by identifying and activating recursively the gene from the rescue set whose deletion contributes the least to the increase of the biomass flux. Gene activation is implemented by restoring the original $(\alpha_j, \beta_j)$ of the corresponding reactions.

**Supplementary Information**

Supplementary Information is available at the *Molecular Systems Biology* website: http://www.nature.com/msb/journal/v4/n1/full/msb20081.html

## Acknowledgements

This work was partially supported by NSF-MRSEC program (DMR-0520513) at the Materials Research Center of Northwestern University (AEM), DOE under contract DE-AC52-06NA25396 (NG), LLNL-LDRD office (06-ERD-061) and DOE under contract W-7405-Eng-48 (EA), and NIH U01 A1070499-01 and 1P20 CA11300-01 (ALB).

**Table I**: Summary of new terminology and effects associated with the identification of genetic rescues interactions.

| Terminology | Definition | Computational method |
|---|---|---|
| Synthetically viable gene pair[(a)] | Removal of one gene is lethal but deletion of a second gene rescues the cell | MOMA and FBA |
| Optimally essential gene[(b)] | Gene deletion leads to zero growth rate in growth-maximizing states | FBA |
| Suboptimally essential gene[(c)] | Gene deletion leads to zero growth rate but growth is possible in optimal states | MOMA and FBA |
| Lazarus effect | Gene deletion restores the growth of otherwise nonviable mutants | MOMA and FBA |
| Suboptimal recovery | Gene deletion increases the growth of already growing strains | MOMA and FBA |

[(a)] Synthetically viable gene *sets* are defined analogously for interactions involving more genes.

[(b)] These genes are *essential* for growth regardless of the state of the other genes.

[(c)] The deletion of these genes is *lethal* but the genes themselves are not *essential*.

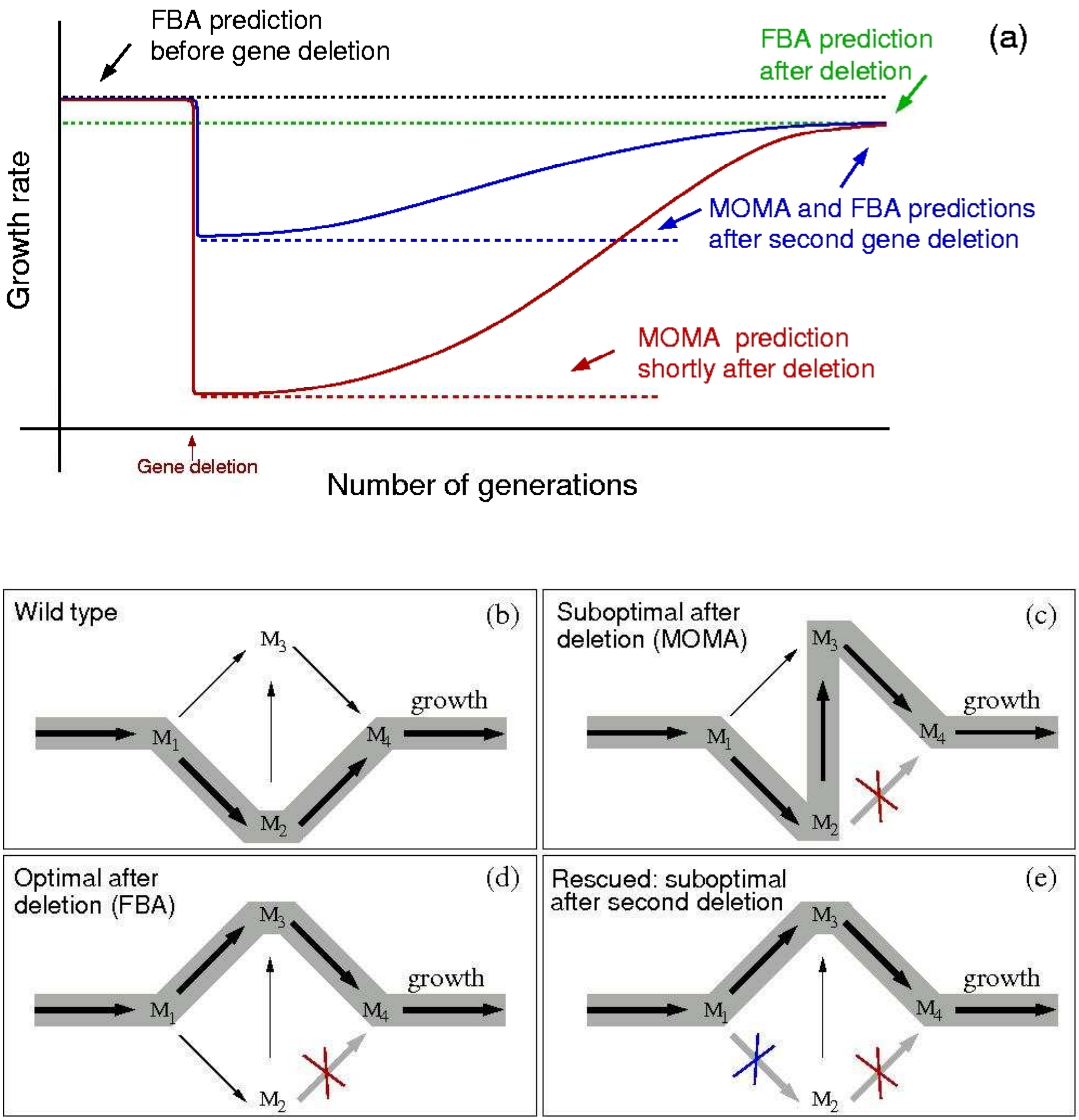

**Figure 1:** Schematic illustration of the consequences of gene deletion on the organism's growth rate. **(a)** The growth rate following the deletion of an enzyme-encoding gene often drops, but after many generations may recover to a new optimal value not very different from the original one (red line). The optimal growth rate before and after the deletion is predicted by FBA (black and green dotted lines). The blue line indicates the predicted buffering effect of additional gene deletions: by deleting appropriately selected additional genes, the suboptimal growth rate shortly after gene deletions is higher than without the rescue deletions. **(b)-(e)** The effect of rescue deletions on the fluxes of a metabolic network, where $M_1$ … $M_4$ represent metabolites and the width of the arrows represents the strength of individual fluxes.

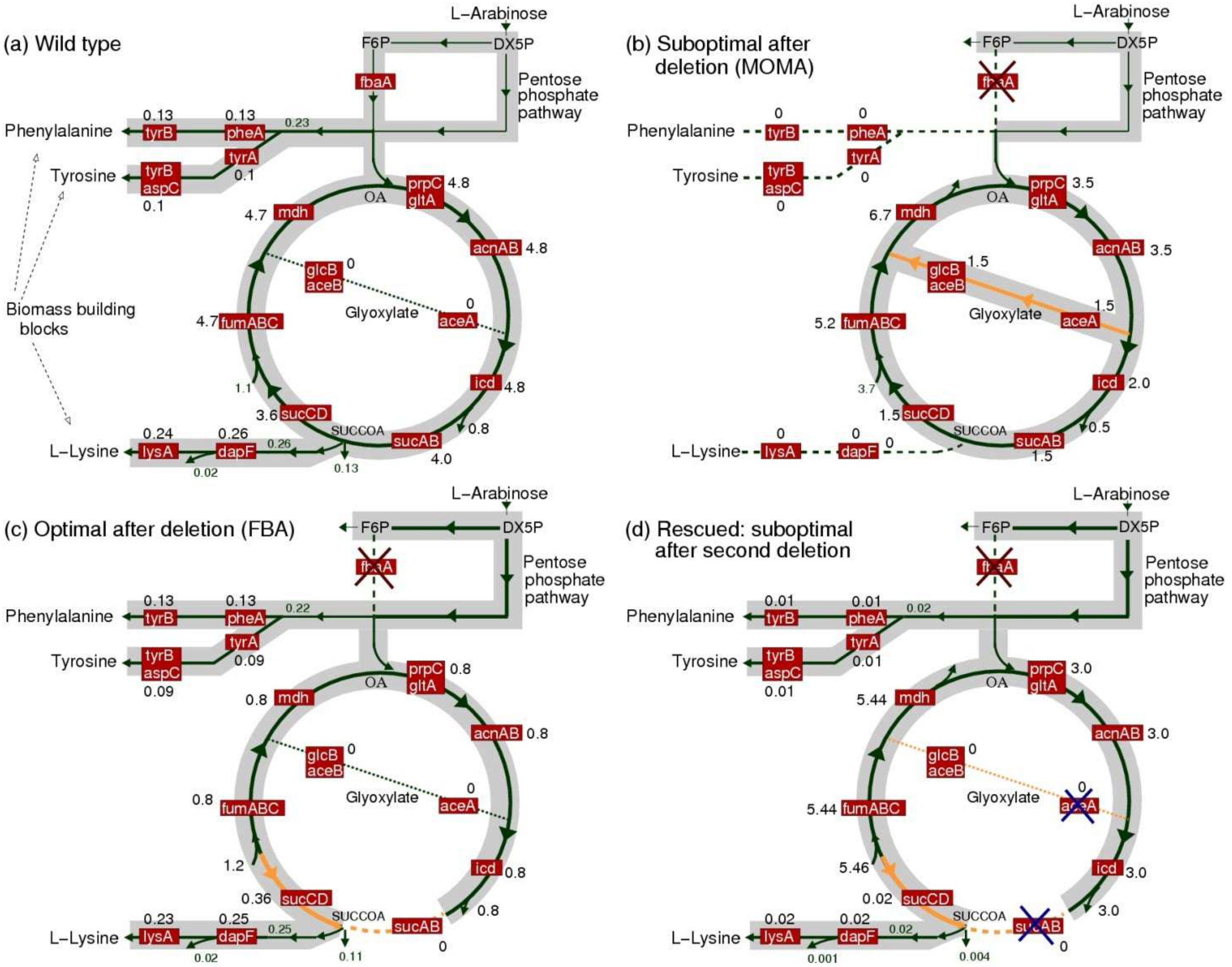

**Figure 2:** Distribution of metabolic fluxes in the *E. coli*'s TCA cycle in arabinose minimal medium for **(a)** wild-type organism predicted by FBA, **(b)** *fbaA*-mutant predicted by MOMA, **(c)** optimal state of *fbaA*-mutant predicted by FBA, and **(d)** *fbaA*-mutant with the rescue deletions of genes *aceA* and *sucAB*, predicted by MOMA. Key flux changes are highlighted in orange. Note that the metabolic flux pattern predicted by MOMA after the *fbaA* deletion (panel (b)) is similar to the wild-type fluxes (panel (a)). With the rescue deletions, however, MOMA predicted fluxes (panel (d)) are brought closer to the FBA predicted fluxes (panel (c)), restoring the organisms' ability to produce biomass. While we show a double deletion for its pedagogical value, we note that the deletion of *aceA* alone is sufficient to rescue the mutant (see Fig. 3(a)) and that the mutant can also be rescued with other single-gene deletions (see Fig. 4(b) and Supplementary Information 4).

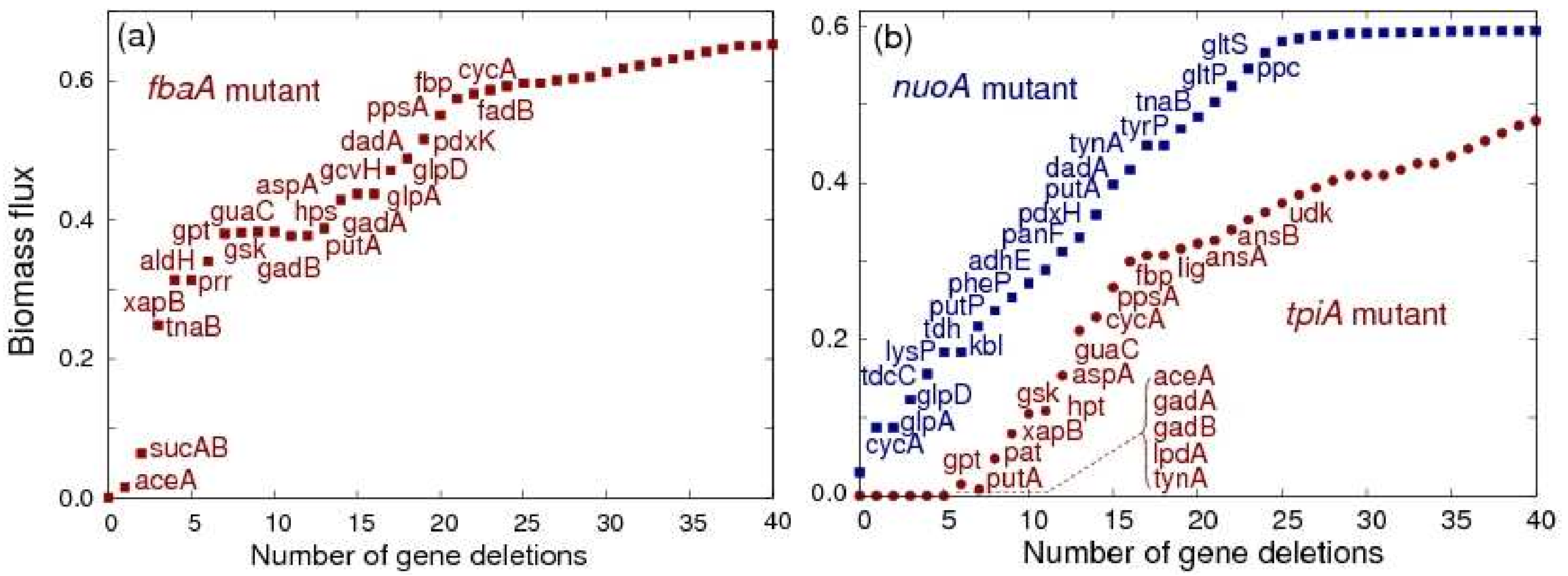

**Figure 3:** The impact of rescue deletions. **(a)** Predicted biomass production for the *fbaA*-mutant of *E. coli* in arabinose minimal medium as a function of the number of rescue deletions when starting with *aceA* and *sucAB*. Deleted rescue genes are indicated in the figure. **(b)** Biomass production of *tpiA*- and *nuoA*-deficient mutants in glucose minimal medium as function of the number of individual rescue deletions. Deleted genes are indicated in the figure. The optimal biomass flux remains unchanged with the addition of rescue deletions. The biomass fluxes are normalized by the wild-type flux $G^{wt}_{FBA}$= 0.745 mmol/g DW-h in (a) and 0.908 mmol/g DW-h in (b).

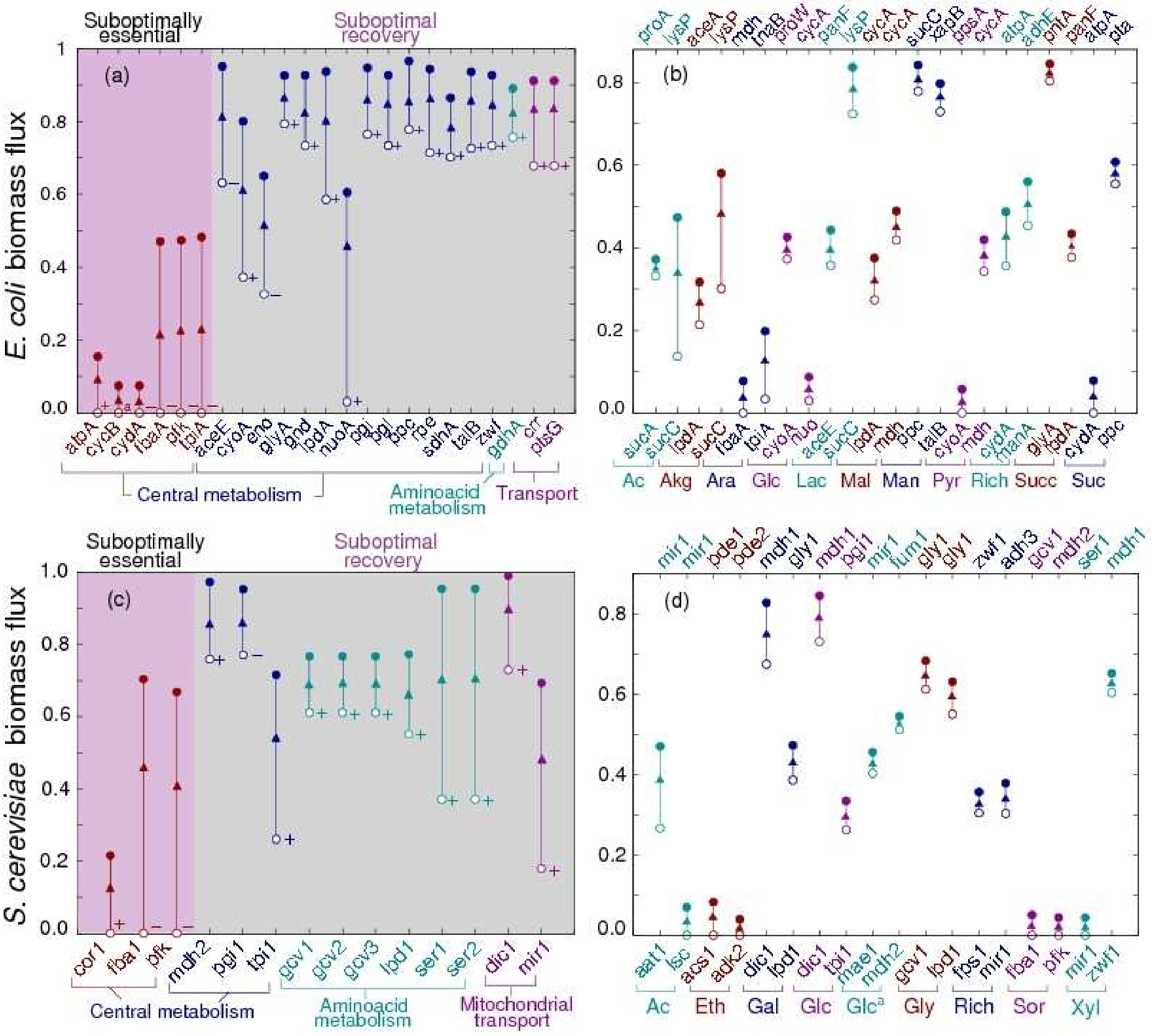

**Figure 4:** The impact of rescue deletions for *E. coli* (a,b) and *S. cerevisiae* (c,d) gene-deficient mutants. **(a,c)** Predicted biomass production before (○) and after (●) rescue deletions in glucose minimal media. The mutants are generated through the deletion of the genes shown at the x-axis. We show the results for all mutants with $G^{1}_{MOMA} < G^{1}_{FBA}$ such that $G^{1}_{MOMA} \le 0.8\ G^{wt}_{FBA}$ and $G^{1}_{FBA} \ge 0.2\ G^{wt}_{FBA}$. If the rescue deletion changes the growth rate from zero to some positive value, we observe the Lazarus effect, applying to suboptimally essential genes (left). If the rescue deletion only enhances the growth rate, we observe a suboptimal recovery (right). The experimental information on the lethality of the original *E. coli* (Edwards and Palsson, 2000; Gerdes *et al*, 2003; Baba *et al*, 2006; PEC, 2007) and *S. cerevisiae* (Giaever *et al*, 2002; Steinmetz *et al*, 2002; SDG, 2007) gene-deficient mutants is indicated with (+) for viable mutants, (-) for nonviable mutants, and ([a]) for a gene absent in the databases. **(b,d)** Same as in (a,c) for single-gene rescue deletions in various media. We show selected mutants with significant biomass improvements after the rescue deletion of a single gene. The rescue deletion is indicated at the top, and the tested media are indicated at the bottom. The abbreviations stand for acetate (Ac), $\alpha$-ketoglutarate (Akg), arabinose (Ara), ethanol (Eth), galactose (Gal), glucose (Glc), glucose anaerobic ($Glc^a$), glycerol (Gly), lactate (Lac), malate (Mal), mannose (Man), pyruvate (Pyr), rich medium (see Supplementary Information 1), sorbitol (Sor), succinate (Succ), sucrose (Suc), and xylose (Xyl). The biomass fluxes are normalized by the wild-type flux $G^{wt}_{FBA}$ in all panels. In units of mmol/g DW-h, the wild-type fluxes for *E. coli* are 0.187 (Ac), 0.535 (Akg), 0.745 (Ara), 0.908 (Glc), 0.367 (Lac), 0.388 (Mal), 0.908 (Man), 0.303 (Pyr), 2.87 (Rich), 0.418 (Succ), and 1.37 (Suc), while for *S. cerevisiae* they are 0.189 (Ac), 0.311 (Eth), 0.703 (Gal), 0.819 (Glc), 0.180 ($Glc^a$), 0.532 (Gly), 1.34 (Rich), 0.798 (Sor), and 0.742 (Xyl). All the genes involved in the rescues of (a) and (c) are listed in Supplementary Information 2 and 3, while the minimum rescue sets are listed in Tables SII and SIII (Supplementary Information 1), respectively. The alternative rescue genes for each media in (b) and (d) are listed along with the corresponding recoveries in Supplementary Information 4 and 5, respectively.

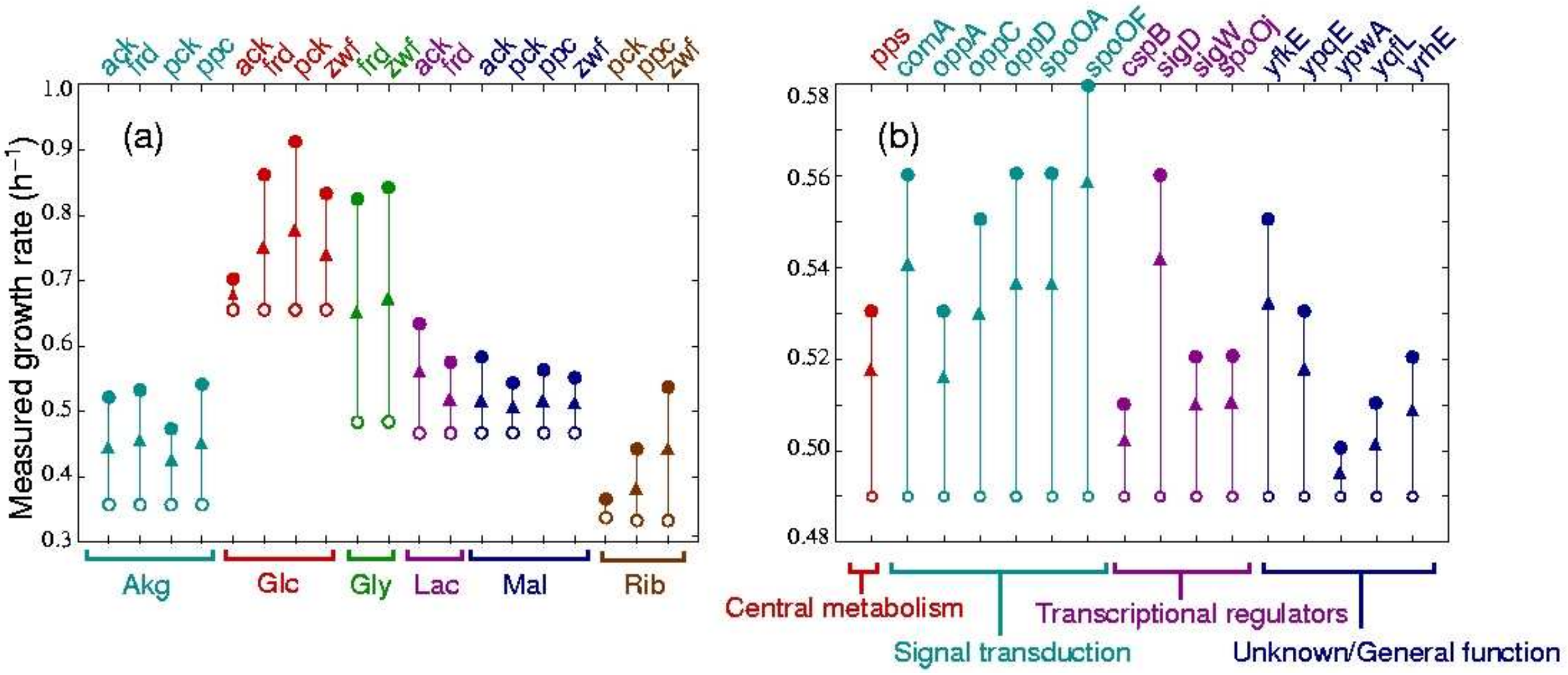

**Figure 5:** Experimental evidence that gene deletions can enhance suboptimal growth rates: growth rate before (○) and after (●) gene deletions for **(a)** *E. coli* MG1655 (Fong and Palsson, 2004) and **(b)** *B. subtilis* 168 (Fischer and Sauer, 2005). The deleted genes are indicated at the top. All genes in panel (a) are involved in the catalysis of central metabolic reactions, and growth is measured after 10 days in $\alpha$-ketoglutarate (Akg), glucose (Glc), glycerol (Gly), lactose (Lac), malate (Mal), and ribose (Rib) media. The carbon source in panel (b) is glucose.